# Fractal Illusions: An Experimental Study of Long-Range Sentence-Length Correlations in Randomly Generated Natural Language Texts

**Ying Zeng**, School of Culture and Communication, Shandong University at Weihai, Shandong, China

**Junying Cui**, Department of Physics, University of Fribourg, Fribourg, Switzerland

**Lejun Li**, Business School, University of Shanghai for Science and Technology, Shanghai, China

**Correspondence:** Ying Zeng, School of Culture and Communication, Shandong University at Weihai, Shandong, China. E-mail: zengying29@163.com

**Abstract**

This study re-evaluates the assumption that long-range correlations in sentence length are a fundamental feature of natural language and a marker of literary style. While previous research has suggested that punctuation marks—particularly full stops—generate structural regularities in narrative texts, our experiments challenge this view. Using Chinese as the primary language, supplemented with English, we constructed randomized linguistic sequences through three distinct methods. Surprisingly, these randomized texts also exhibit long-range correlations in sentence length, some even with stronger fractal characteristics than those found in canonical literary works. These findings suggest that the presence of long-range correlations in sentence length is not sufficient to indicate authorial intention, structural depth, or literary value. We argue that punctuation-induced long-range correlations have limited value for literary stylistic analysis and cannot be considered a fundamental characteristic of human writing.

**Keywords**: randomized texts, sentence length, punctuation and structure, literary style, fractal analysis

# 1. Introduction

A wide range of natural phenomena that appear "random" actually exhibit self-similar fractal patterns, such as the crystalline forms of snowflakes and the jagged edges of clouds and coastlines. It is, therefore, striking to scholars in the humanities that human language, much like these natural phenomena, also contains implicit fractal structures.

Methods originating from statistical physics, particularly time-series analysis, have been increasingly applied to literary texts to detect such fractal patterns. They can primarily be observed through three key indicators: the Hurst exponent (H), the power-law exponent (β), and the multifractal spectrum width (Δα). The Hurst exponent (H), ranging from 0 to 1, quantifies the long-range dependence of a time series, with H > 0.5 indicating persistence, H < 0.5 indicating anti-persistence, and H = 0.5 suggesting no correlation. This H value can be corroborated by the power-law exponent, β, which reflects the frequency of a pattern's occurrence. A higher β suggests stronger periodicity, while a lower β indicates a lower frequency. The Δα value is used to distinguish between monofractal and multifractal structures.

While it is understood that long-range correlations at the letter and word levels may arise from semantic clustering—'groups of words associated by affine semantic hierarchies form a complex arrangement of cohesive clusters even over spans of entire corpora, thereby giving rise to long-range order in human written records' (Montemurro and Pury, 2002)—the origins of analogous correlations in sentence length remain less clear. Unlike word choice, which offers relatively limited flexibility, writers have much greater freedom in determining sentence length. From a strictly linguistic perspective, nearly any long sentence can be split and restructured into shorter ones, just as any short sentence can be expanded. This inherent freedom gives writers near-complete control over sentence length.

One of the most representative studies on long-range correlations in sentence length is by Drożdż et al. (2016). By examining sentence length variability (SLV) in numerous well-known literary works, they discovered that periods play a distinctive role in these correlations. They proposed that this finding could serve both as an indicator of factors shaping human language and as a tool for stylistic measurement. This study will delve deeper into their hypothesis, critically evaluating whether these correlations have a meaningful impact on literary style and whether fractality is truly an inherent characteristic of language or style.

Given that human-authored natural language texts typically possess logical and semantic coherence, this study adopts a different approach. We use several experimental methods to generate randomized texts that intentionally disrupt narrative logic and semantic coherence, rendering them semantically incoherent and unreadable. Our randomization methods fall into three categories: (1) randomly inserting sentence-final and major punctuation marks into texts, (2) generating text by randomly striking keys on a computer keyboard, and (3) randomly combining sentences selected from different sources. These randomized corpora are then compared with coherent texts to examine whether sentence-final punctuation, such as periods, still triggers long-range correlations, as measured by three key indicators: the Hurst exponent (H), the power-law exponent ($\beta$), and the multifractal spectrum width ($\Delta\alpha$).

## 2. Materials and Methods

*2.1 Experimental Methods and Data*

As the majority of our randomized texts were generated through the participation of native Chinese speakers, the primary materials for this study are in Chinese, with a smaller subset in English.

The materials are categorized into three types, each derived from one of the following three experimental methods.

*2.1.1 Experimental Method 1*

Twenty Chinese university students were assigned the task of randomly inserting punctuation into a given text. The method was divided into two approaches: Approach A and Approach B.

**Approach A**: The original Chinese novel *Fan Hua* had all punctuation removed and was then divided into twenty equal parts, each assigned to one of twenty Chinese university students. Participants were instructed to randomly insert sentence-final and major punctuation marks into their assigned sections. The modified fragments were then compiled into a single text, henceforth referred to as *Hua*.[1] The text contained 13,292 sentences, determined by terminal punctuation marks.

A comparison of the text segments between *Hua* and the original *Fan Hua* is shown in Table 1.

Table 1. Comparison of the text segments between *Hua* and the original *Fan Hua*

| *Fan Hua* (original text) | *Fan Hua* (English Translation) | *Hua* (experimental text) | *Hua* (English Translation) |
|---|---|---|---|
| 阿宝说，我是阿宝。女声说，我雪芝呀。阿宝嗯了一声，回忆涌上心头。阿宝低声说，现在不方便，再讲好吧，再联系。阿宝挂了电话。夜风凉爽，两人闷头走路，听见一家超市里，传来黄安悠扬的歌声， | Abao said, "It's me, Abao." A woman's voice replied, "It's Xuezhi." Abao gave a brief "mm" as a rush of memories came flooding back. In a low voice, he said, "It's not convenient right now. Let's talk later—I'll get in touch." He hung up the phone. The two of them walked on in silence, heads down, while a cool night breeze brushed past. From a nearby supermarket came the gentle, melodic voice of Huang An, drifting through the air. | 喂阿宝说我是阿，宝女声说我雪，芝呀阿宝嗯了一声回。忆涌上心头阿，宝低声说现，在不方便，再，讲好吧再联，系阿宝挂了电，话夜风凉，爽两人闷，头走路听见一家超，市里传来，黄安悠，扬的歌声。 | Hello A, bao said I'm A, bao. A woman's voice said I'm Xue, zhi yeah. A, bao hmm'd a bit, and mem. ories surged to his heart. A, bao said softly no, w's not conve, nient, let's tal, k later, get in to, uch. A, bao hung up the ph, one. The night breeze was co, ol, the two walked withhea, ds down, heard from a su, permarket the melo, dious voice of Huang An.[2] |

From the experimental text, it is evident that the random insertion of punctuation marks disrupted the semantic coherence of the original text, producing numerous sentences that are fragmented and incomprehensible.

**Approach B**: To further minimize any potential influence of the original text on participants' punctuation choices, the character order of *Fan Hua* was completely randomized before the experiment. The scrambled text was divided into twenty equal parts, each assigned to one of the twenty Chinese university students. As in Approach A, participants were instructed to insert sentence-final and major punctuation marks at random positions within their assigned sections. The modified fragments were subsequently assembled into a single text, henceforth referred to as *Fan*. The text contains 7,908 sentences, determined by terminal punctuation marks.

A comparison of the text segments between *Fan* and the original *Fan Hua* is shown in Table 2.

Table 2. Comparison of the opening segments between *Fan* and the original *Fan Hua*

| *Fan Hua* (original text) | *Fan Hua* (English Translation) | *Fan* (experimental text) | *Fan* (English Translation) |
|---|---|---|---|
| 沪生经过静安寺菜场，听见有人招呼，沪生一看，是陶陶，前女朋友梅瑞的邻居。 | Passing by the Jing'an Temple market, Husheng heard someone calling out to him. He looked over—it was Taotao, a neighbor of his ex-girlfriend Meirui. | 吓间大闷子米张芬人，面讲生作乐门姆，话上果小超点珍得熟。沪进妇妈说沪果，解到银码腿边连钳， | Fright interval big stifler rice Zhang Fen person, face talk life make music door mu, speech up fruit small super spot Zhen get ripe. Hu enter woman mom say Hu fruit, understand arrive silver code leg side link pliers, |

In contrast to the partially garbled sentences in the previous experimental text, *Hua*, every sentence in *Fan* consists entirely of scrambled characters, rendering it completely unreadable.

### 2.1.2 Experimental Method 2

One hundred Chinese university students were asked to type on a computer keyboard randomly.[3] The texts typed by each participant were then assembled into a single text, named *Jing*, with a total of over 300,000 characters. The text contained 4,116 sentences, as counted by terminal punctuation marks.

Below is an excerpt from *Jing*:

词不达意和基础宣汉欣慰发挥切换殿下是粗啊的女你是荆钗布裙独处才是瓜的回复回去独特上次问别人回复个商场"新闻席梦思床老妈大客户大哥 vu 啊的内部温度为白云苍狗松村谦三觉得怒啊的鬼天气是 v 粗纹产生传播 XH 吧很符文言出法随要吃你那恶臭金翅鸟无处不问符文孵化器恶意风格我依然句吧是一个呀都不窜看心情现场可是 M 想冒充，

（Words not reach meaning and foundation declare Han pleased exert switch Your Highness is rough ah female you are hairpin cloth skirt alone is melon's reply go back unique last time ask others reply a mall 'news Simmons bed mom big client big bro vu ah internal temperature is white cloud blue dog Matsumura Kenzō thinks angry ah ghost weather is v rough grain produce transmit XH ba very rune literary issue method follow want eat you that foul golden-wing bird nowhere not ask rune incubator malicious style I still sentence ba is a one ya all not jump see mood live scene but M wants to impersonate,）

We can observe that in this randomly typed text, punctuation marks are sparse, often appearing only after many characters. As a result, the sentences become very long, containing recognizable words and phrases but overall forming an incoherent jumble of characters.

### 2.1.3 Experimental Method 3

150 university students were asked to randomly select forty books and then pick one sentence from each book (using sentence-final punctuation as the marker). These selected sentences were then assembled into a single text of 6061 sentences, named *Cao Ji*.

Below is an excerpt from *Cao Ji*:

①家丽沉吟，一会儿，才让老五老六起来，对姊妹几个说："咱们家遭了一个难，谁也别说了，爸妈回来，就说是打雷劈着了树着的火。"②照例每年夏天有一批中国留学生学成回国。③有一阵风吹过来，把颂莲的裙子吹得如同飞鸟，颂莲这时感到一种坚硬的凉意，像石头一样慢慢敲她的身体，颂莲开始往回走，往回走的速度很快，回到南厢房的廊下，她吐出一口气，回头又看那个紫藤架，架上倏地落下两三串花，很突然地落下来，颂莲觉得这也很奇怪。[4]

(① Jia Li pondered for a moment before allowing Fifth and Sixth to get up, then told her sisters, "Our family has suffered a disaster. Let no one speak of it. When Mom and Dad return, just say it was lightning that struck the tree and started the fire." ② As usual, every summer a group of Chinese students returned home after completing their studies abroad. ③ A gust of wind blew over, making Songlian's skirt billow like a bird in flight. Songlian then felt a hard, cold sensation, slowly tapping her body like a stone. Songlian began to walk back, her pace quickening. When she returned to the corridor of the south wing, she let out a breath and looked back at the wisteria trellis. Suddenly, two or three clusters of flowers fell from the trellis, very unexpectedly. Songlian found this very strange as well.)

The three sentences in the above excerpt come from different authors. In *Cao Ji*, each sentence is either from a different author, from different works by the same author, or from different sentences within the same work.

We translated the third experimental text, *Cao Ji*, into English, then constructed a sentence length sequence and evaluated these parameters.

Additionally, we included two representative works of shuffle literature (Husárová & Montfort, 2012): the Chinese translation of Marc Saporta's novel Compos*ition No. 1,* and the original English version of B.S. Johnson's *The Unfortunates*, evaluating both their original and randomized versions by measuring the same set of indicators.

*2.2 Processing of Sentence Length Data*

Sentence length in the texts is processed in two distinct ways. The first method constructs a time series of character counts (for Chinese) or word counts (for English) between two sentence-final punctuation marks that signal the completion of a thought. These marks include the period, question mark, exclamation mark, and ellipsis. The second method constructs a time series between major pause-indicating punctuation marks, which, in addition to sentence-final punctuation, also comprise the comma, colon, and semicolon. All punctuation marks—regardless of type—are systematically mapped to discrete symbolic values. The time series is then constructed according to the order in which these symbols occur in the original text.

Given the ideographic nature of Chinese, the fundamental unit for constructing the sentence sequence can be either a character or a word. Analysis of multiple Chinese works indicates that both character-based and word-based measurements yield clear fractal patterns, with character-based measurements producing stronger fractal characteristics. Accordingly, in this study, Chinese texts are primarily analyzed with characters as the basic unit of the time series, while English texts are analyzed with words as the basic unit.

*2.3 Methods: Multifractal Detrended Fluctuation Analysis and Power Spectral Analysis*

This study integrates Multifractal Detrended Fluctuation Analysis (MFDFA) with Power Spectral Analysis. MFDFA, first proposed by Kantelhardt et al. (2002), is a widely used method for characterizing the multifractal properties of non-stationary time series, and is particularly suitable for identifying nonlinear, multiscale, and nonstationary structures in linguistic data. Prior to power spectral analysis, all sequences were centered and standardized (mean = 0, standard deviation = 1), while the original scale data were retained for MFDFA.

*2.3.1 MFDFA Procedure*

Given a one-dimensional time series $x(i), i = 1,2, \ldots, N$, the multifractal analysis procedure is as follows:

Step 1: Cumulative deviation sequence (trajectory)

$$Y(i) = \sum_{k=1}^{i} [x(k) - \bar{x}], i = 1,2, \ldots, N$$

where $\bar{x}$ is the mean of the sequence.

Step 2: Interval partitioning and detrending

The cumulative sum series $Y(i)$ is divided into $N_s = \lfloor N/s \rfloor$ non-overlapping segments of equal length $s$. Within each segment $v$, a polynomial function $P_v^{(m)}(i)$ of order $m$ is fitted to the local profile to capture the underlying trend. In this study, we adopt a second-order polynomial ($m = 2$). The variance of the detrended residuals in segment $v$ is then computed as:

$$F^2(v, s) = \frac{1}{s} \sum_{i=1}^{s} \{Y[(v-1)s + i] - P_v^{(m)}(i)\}^2$$

Step 3: Calculation of q-order fluctuation function

For each scale $s$, the detrended residuals across all segments are normalized by the q-th order and aggregated to obtain the fluctuation function:

$$F_q(s) = \left\{ \frac{1}{2N_s} \sum_{v=1}^{2N_s} [F^2(v, s)]^{q/2} \right\}^{1/q}, q \neq 0$$

In this study, to avoid divergence and numerical instability associated with $q = 0$, only $q \in [-4,4] \setminus \{0\}$ was considered, resulting in a total of 40 q-values. The case $q = 0$ is excluded because it involves logarithmic averaging, which is unstable and not suitable for standard normalization.

The scale parameter $s$ is logarithmically spaced, with approximately 100 values sampled from the interval $s \in [3,1000]$. If the maximum scale $s_{max} = 1000$ exceeds the series length, it is automatically adjusted to $s_{max} = N/5$ to ensure accurate computation. Therefore, the effective scales used satisfy $s < N/5$.

To guarantee statistical reliability at each scale, it is required that the series length satisfy $N \geq 5s_{max}$; otherwise, the maximum scale $s_{max}$ will be further reduced accordingly.

Step 4: Generalized Hurst exponent extraction

For different scales $s$, a log–log linear fit is applied between $F_q(s)$ and $s$. If a power-law relationship is observed:

$$F_q(s) \sim s^{h(q)}$$

then $h(q)$ is defined as the generalized Hurst exponent. To reduce small-scale noise and large-scale boundary effects, the fitting interval is selected from the 20th point to the maximum scale, i.e., $s \in [s_{20}, N/5]$. A linear regression is performed within this interval.

If $\log F_q(s)$ and $\log s$ are linearly correlated, then the time series exhibits scale invariance and multifractal properties at the corresponding $q$-order.

Step 5: Singularity spectrum calculation

The singularity exponent $\alpha$ and the singularity spectrum $f(\alpha)$ are obtained through the Legendre transform:

$$\alpha = \frac{d}{dq}[qh(q) - 1], f(\alpha) = q\alpha - [qh(q) - 1]$$

The singularity spectrum $f(\alpha)$ characterizes the distribution and complexity of fluctuations with different intensities in the time series.

Step 6: Spectrum Width Index $\Delta\alpha$

Here, $\Delta\alpha$ represents the width of the singularity spectrum, and a larger value indicates a richer and more diverse multifractal structure.

The multifractal strength is measured by the spectrum width $\Delta\alpha = \alpha_{max} - \alpha_{min}$. Generally:

- $\Delta\alpha \leq 0.1$: Very weak multifractality or monofractal
- $0.1 < \Delta\alpha \leq 0.2$: Weak multifractality
- $\Delta\alpha > 0.2$: Strong multifractality

The left side of the spectrum (corresponding to $q > 0$) emphasizes large fluctuations, while the right side ($q < 0$) highlights minor fluctuations. A longer left tail indicates dominance of large-scale fluctuations, whereas a longer right tail reflects dominance of fine-scale variations.

*2.3.2 Power Spectral Density Analysis*

To further examine the long-range correlation and linear memory properties of the sequence, we introduce Power Spectral Density (PSD) analysis as a complementary approach to MFDFA. Specifically, we apply the Fast Fourier Transform (FFT) to the centered and standardized sequence $x(i)$, and compute the power spectral density $S(f)$:

$$\mathcal{F}[x(i)] = \hat{x}(f), S(f) = \frac{|\hat{x}(f)|^2}{N}$$

Here, $f$ denotes the normalized frequency. To minimize edge effects and improve spectral resolution, the sequence is zero-padded to the nearest $2^k$ length before transformation. The frequency unit is given by $f = k/N$, where $k$ is the frequency index.

In practice, the power-law behavior of the spectrum is examined in the low-frequency range that satisfies $\log(f) < -2.75$. This restriction helps ensure linear scaling and avoids distortion from local fluctuations or noise. A linear regression is performed on the log-log plot of $S(f)$ versus $f$ in this range to estimate the spectral exponent $\beta$, defined via:

$$S(f) \sim f^{-\beta}$$

The spectral exponent $\beta$ characterizes the memory and complexity of the sequence:

- $\beta = 0$: White noise (no correlation)
- $0 < \beta < 1$: Long-range positive correlation, the β value usually ranges between ¼ (or 0.25) and ¾ (or 0.75) (Drożdż *et al*. 2016)
- $\beta = 1$: 1/f noise (pink noise)
- $\beta > 1$: Strong dependence or non-stationarity
- $\beta < 0$: Anti-persistent behavior

Theoretically, the spectral exponent $\beta$ and the Hurst exponent $H = h(2)$ obtained from MFDFA satisfy the following relationship:

$$\beta = 2H - 1$$

The Hurst exponent $H$ reflects the degree of long-range dependence in the sequence. When $H > 0.5$, the sequence tends to exhibit persistent (inertial) behavior, implying that future values are likely to follow past trends. In contrast, $H < 0.5$ suggests anti-persistence, where increases are more likely to be followed by decreases, and vice versa. When $H \approx 0.5$, the series behaves similarly to white noise, indicating a lack of long-term correlation.

MFDFA and PSD provide complementary perspectives: MFDFA emphasizes local non-stationarities and scale-dependent fluctuations in the time domain, while PSD captures global correlations and scale invariance in the frequency domain. Together, they offer a more comprehensive understanding of the multiscale structures present in the sequence.

## 3. Results and discussion

*3.1 Results of the First Experiment*

**Method A:** From Fig. 1, it is clear that there are pronounced differences in the sentence length sequences between *Hua*, with randomly inserted punctuation, and the original text, *Fan Hua*. Compared with *Fan Hua*, *Hua* exhibits sharper peaks, reflecting the presence of exceptionally long sentences, along with denser clusters of short sentences in certain sections.

As shown in Table 3, [5]*Hua* yields higher Hurst exponents for both punctuation schemes ($H_1$ = 0.81, $H_2$ = 0.74) than *Fan Hua* (0.65, 0.64), indicating stronger long-range correlations regardless of segmentation type. The corresponding spectral exponents $\beta_1$ and $\beta_2$ (0.71, 0.64) are also markedly higher than those of *Fan Hua* (0.35, 0.32), confirming this trend in the frequency domain. The multifractal spectrum widths ($\Delta\alpha_1$ = 0.33, $\Delta\alpha_2$ = 0.14) are comparable to those of *Fan Hua* (0.37, 0.18), suggesting that the degree of multifractality is largely unaffected by the random insertion of punctuation.

Table 3. Comparison between *Hua* and *Fan Hua*

| Text | $H_1$ | $H_2$ | $\beta_1$ | $\beta_2$ | $\Delta\alpha_1$ | $\Delta\alpha_2$ |
| --- | --- | --- | --- | --- | --- | --- |
| *Hua* | 0.81 | 0.74 | 0.71 | 0.64 | 0.33 | 0.14 |
| *Fan Hua* | 0.65 | 0.64 | 0.35 | 0.32 | 0.37 | 0.18 |

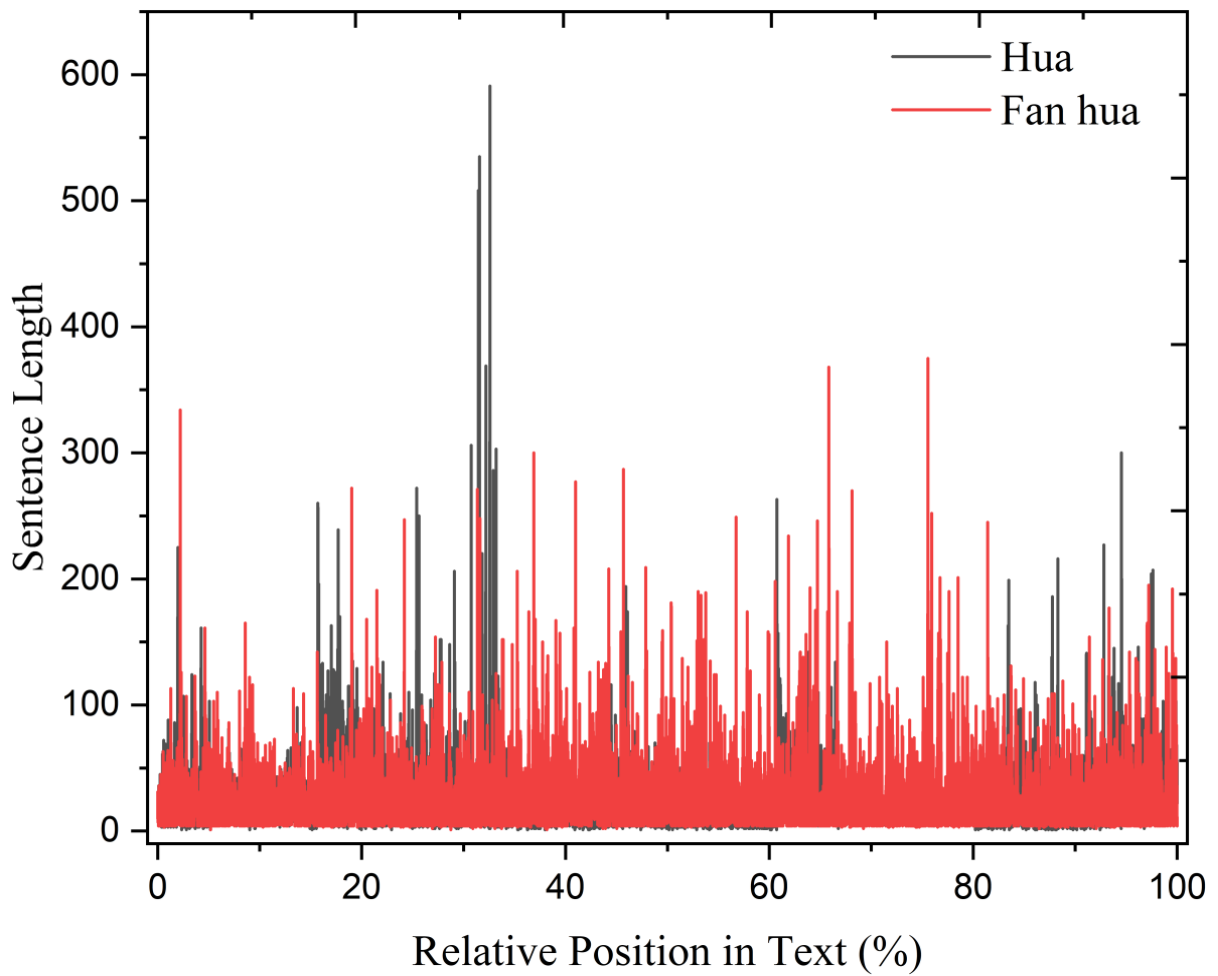

**Figure 1.** Sentence length sequence comparison between *Hua* (black) and the original *Fan Hua* (red).

**Method B:** As shown in Fig. 2, the sentence-length peaks in *Fan* are sharper than those in *Fan Hua*—and also sharper than those in the previous experimental text *Hua*—with the longest sentence exceeding 3,000 characters, which alters sentence boundaries and changes the distribution and variability of sentence lengths. Table 4 shows that *Fan* exhibits stronger fractal parameter values: $H_1 = 0.82$ and $H_2 = 0.89$, both significantly higher than those of *Fan Hua* (0.65, 0.64). Likewise, $\beta_1$ and $\beta_2$ (0.74, 0.72) are markedly greater than those of *Fan Hua* (0.35, 0.32), indicating stronger long-range correlations in the frequency domain. The most notable difference is in the $\Delta\alpha$ values: *Fan*'s $\Delta\alpha_1 = 0.88$ and $\Delta\alpha_2 = 0.49$ are approximately two to two-and-a-half times those of *Fan Hua*, indicating a marked increase in the heterogeneity and complexity of sentence-length variation.

Table 4. Comparison between *Fan* and *Fan Hua*

| Text | $H_1$ | $H_2$ | $\beta_1$ | $\beta_2$ | $\Delta\alpha_1$ | $\Delta\alpha_2$ |
|---|---|---|---|---|---|---|
| *Fan* | 0.82 | 0.89 | 0.74 | 0.72 | 0.88 | 0.49 |
| *Fan Hua* | 0.65 | 0.64 | 0.35 | 0.32 | 0.37 | 0.18 |

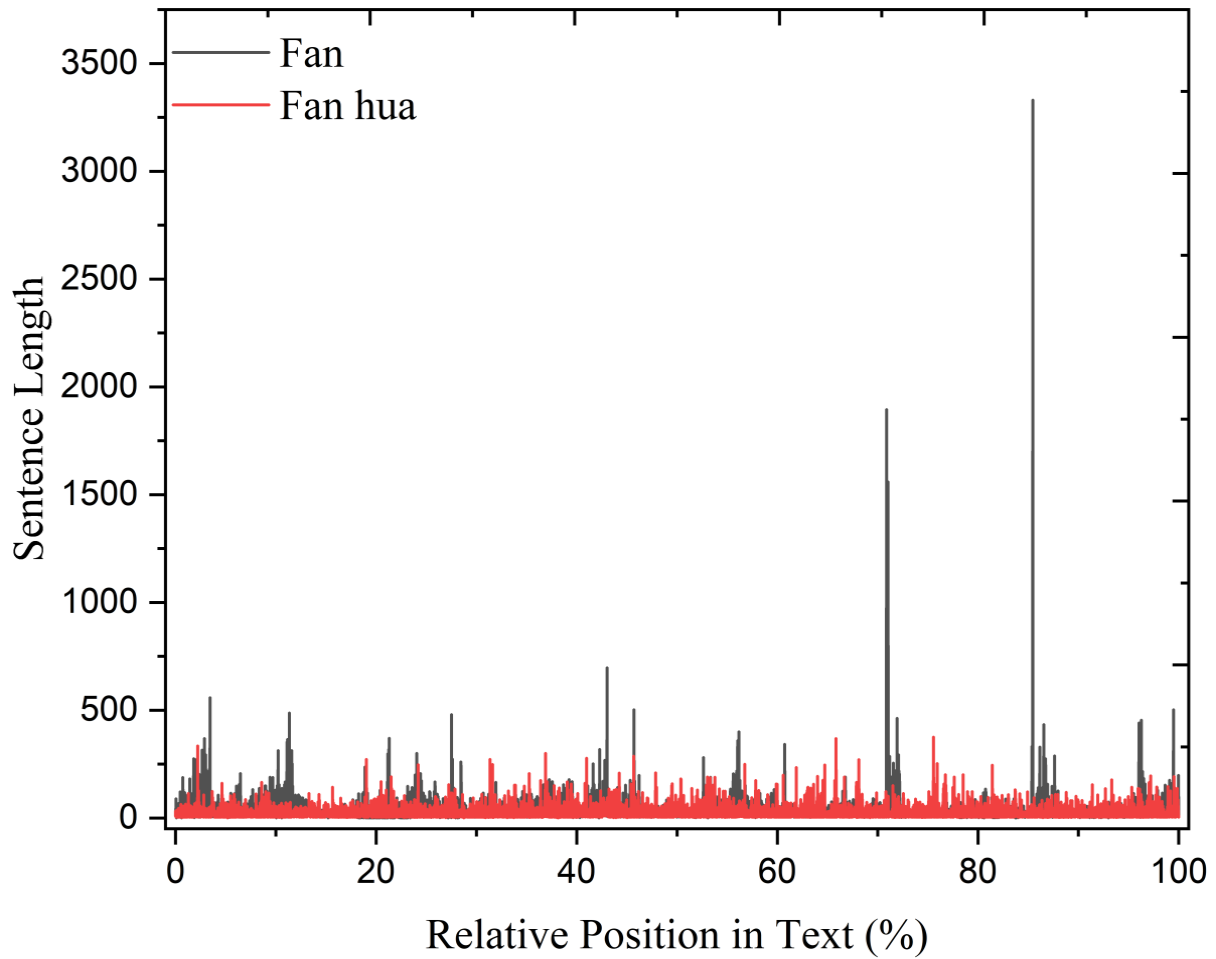

**Figure 2.** Sentence length sequence comparison between *Fan* (black) and the original *Fan Hua* (red).

*3.2 Results of the Second Experiment*

From Fig. 3, it is evident that *Jing* contains extremely long sentences, some of which are close to ten thousand characters in length. As shown in Table 5, the Hurst exponent based on sentence-final punctuation is moderate ($H_1 = 0.54$), whereas the exponent based on major punctuation is notably higher ($H_2 = 0.71$), indicating stronger correlations at the major-punctuation scale. The spectral exponents ($β_1 = 0.35$, $β_2 = 0.48$) are consistent with these observations. The most striking feature is the multifractal spectrum width—$Δα_1 = 1.58$ and $Δα_2 = 0.93$—among the largest observed in this study, indicating exceptionally high structural complexity. Such a result represents a characteristic feature of strong multifractal behavior.

Table 5. The values for *Jing*

| Text | $H_1$ | $H_2$ | $\beta_1$ | $\beta_2$ | $\Delta\alpha_1$ | $\Delta\alpha_2$ |
|---|---|---|---|---|---|---|
| *Jing* | 0.54 | 0.71 | 0.35 | 0.48 | 1.58 | 0.93 |

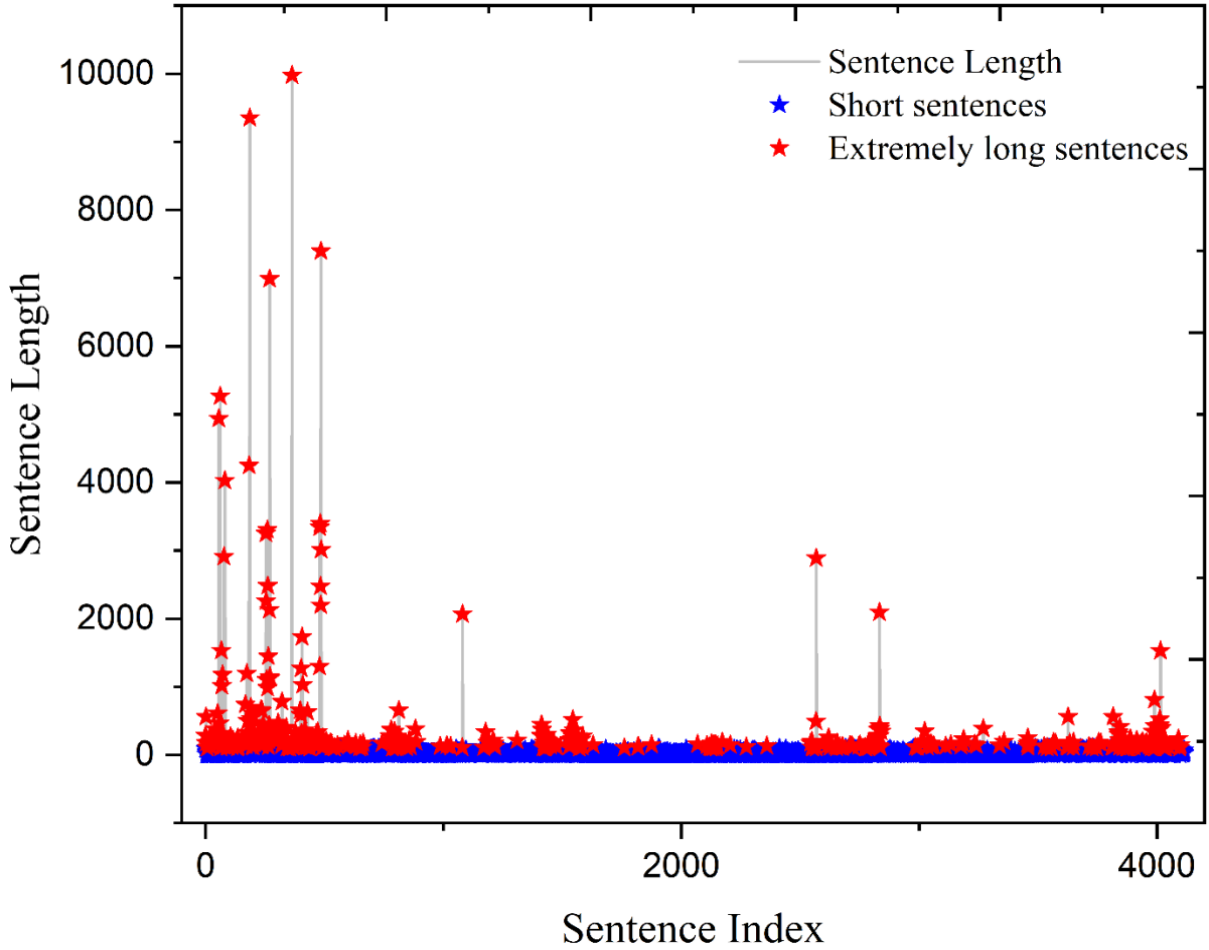

**Figure 3.** Sentence length distribution in *Jing* with detected short and extremely long sentences.

*3.3 Results of the Third Experiment*

As reported in Table 6, the Chinese original of *Cao Ji* exhibits high Hurst exponents ($H_1$ = 0.83, $H_2$ = 0.74) and corresponding spectral exponents ($\beta_1$ = 0.73, $\beta_2$ = 0.44), indicating clear long-range dependence at both punctuation scales. The relatively narrow multifractal spectrum widths ($\Delta\alpha_1$ = 0.21, $\Delta\alpha_2$ = 0.14) suggest moderate diversity in scaling behavior, implying a stable sentence-length organization with limited structural variability.

The accompanying multiscale analysis further supports these observations. Fig. 4(a) presents the raw sentence-length sequence, showing relatively uniform fluctuations without any extremely long sentences, as also evident in Fig. 5, which reflects a stable sentence structure. Fig. 4(b) illustrates the detrended fluctuation function F_q(s), displaying consistent power-law growth. Fig. 4(c) shows the singularity spectrum f(α), with a spectrum width of Δα = 0.208—indicating weak multifractality. Fig. 4(d) presents the power spectrum, exhibiting clear power-law decay with a spectral exponent β = 0.728, corresponding to a Hurst exponent H = 0.834.

Table 6. The values for *Cao ji*

| Text | $H_1$ | $H_2$ | $β_1$ | $β_2$ | $Δα_1$ | $Δα_2$ |
|---|---|---|---|---|---|---|
| *Cao ji* | 0.83 | 0.74 | 0.73 | 0.44 | 0.21 | 0.14 |

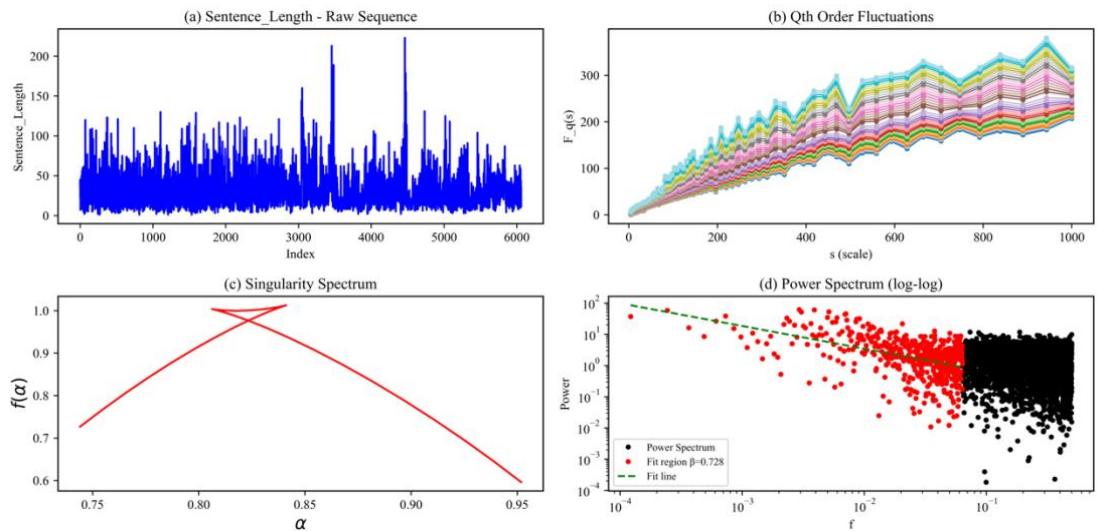

Figure 4. Multiscale analysis of the sentence-length sequence of *Cao Ji*

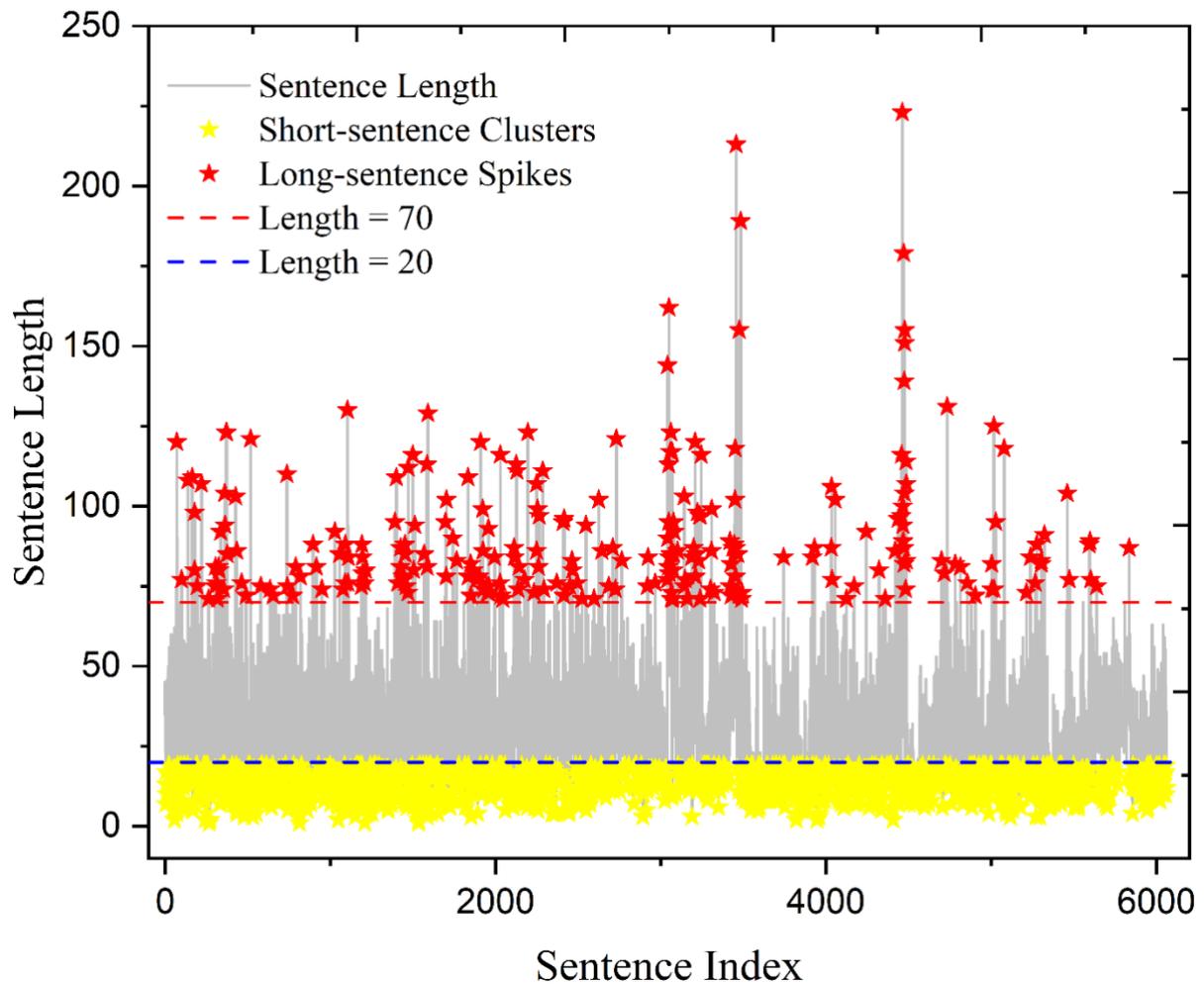

**Figure 5.** Sentence length distribution of *Cao Ji,* with detected short-sentence clusters and long-sentence spikes.

As shown in Table 7, although the parameter values for the English translation of *Cao Ji* are lower than those of the Chinese original, the text still exhibits a clear power-law long-range temporal correlation. All H values are above 0.5, $β_1$ values exceed 0.25, and $β_2$ values are relatively low at 0.11.[6]

Table 7. The values for the English version of *Cao Ji*

| Text | $H_1$ | $H_2$ | $β_1$ | $β_2$ | $Δα_1$ | $Δα_2$ |
|---|---|---|---|---|---|---|
| English version of *Cao Ji* | 0.72 | 0.57 | 0.45 | 0.11 | 0.06 | 0.07 |

Additionally, the results of the two shuffle literature works will be discussed in the next subsection.

*3.4 Overall Discussion of All Experimental Data*

In the first three experimental texts (*Hua*, *Fan*, and *Jing*), punctuation marks such as periods were randomly inserted, resulting in fragmented or even nonsensical sentences. Despite this semantic disruption, the long-range correlation in sentence-length sequences was clearly present and persisted. These meaningless texts still exhibited fractal structures, which suggests that such correlations are not strictly dependent on the original literary structure or the author's deliberate narrative design. This finding challenges the validity of using long-range sentence-length correlations as a measure of literary style.

The results from these experiments appear to validate our initial hypothesis: similar to heart rate variability or speech patterns, long-range correlations in sentence length might be tied to human physiological rhythms, such as breathing, representing a feature of the "natural voice" (Drożdż et al., 2016). Therefore, even when punctuation is randomized and semantic meaning is lost, the sentence structures may still reflect an underlying human marking behavior. This idea is supported by Schenkel et al. (1993), who conducted an experiment to determine if human-generated random numbers exhibited inherent correlations. A participant, tasked with mimicking a random number generator by writing digits from 0 to 9, produced a sequence that retained long-range correlations even after being converted into a binary bitstream.

However, the results of the third experimental text, *Cao Ji*, strongly challenge—and effectively rule out—the hypothesis that long-range correlations in sentence length originate from human physiological rhythms. Although the punctuation marks in *Cao Ji* likewise stem from human behavior, each sentence was randomly selected from different texts written by different individuals. This procedure produced a genuinely random combination of punctuation patterns. The prominent long-range correlations observed in this shuffled text clearly indicate that such correlations cannot be attributed to the physiological rhythm of a single individual.

Dec et al. (2024) studied the novel *Hopscotch*, published in 1963 by the Latin American writer Julio Cortázar. The unique feature of this novel is that readers can skip around and read the chapters out of the printed sequence. The study concluded that at scales larger than the average number of sentences per chapter, no significant temporal correlations were introduced. Each chapter constituted a highly independent block, and its position in the narrative could be arbitrarily changed without altering the statistical properties of SLV.

Similarly, the well-known shuffle literature author B.S. Johnson, who also wrote in the 1960s, created the book *The Unfortunates*. In this work, the chapters were placed in a box, and all chapters were separately bound. With the exception of the first and last chapters, readers could read them in any random order. A power-law long-range temporal correlation also existed in the SLV of both its printed sequence and a randomized chapter version (for details, see Appendix A).

Going further than Cortázar and Johnson was the French writer Marc Saporta, often called the pioneer of shuffle literature. He gave independence to each page (below the chapter level but above the paragraph level), creating a card-based novel called *Composition No. 1*. It consists of over 150 unbound loose-leaf pages, and readers can shuffle the pages like a deck of cards, getting a new narrative with each shuffle. However, in *Composition No. 1*, a power-law long-range temporal correlation also existed in the SLV for both the printed order and a randomized page version. The fractal quality was even better after randomization (see Appendix A).

Our experimental text, *Cao Ji*, shares a similar principle with a work of shuffle literature,[7] but it takes the concept a step further by applying the shuffle principle at a smaller scale than the chapter and page levels of the three works mentioned above, as it combines each sentence randomly. The significant power-law long-range temporal correlations that still appeared in the SLV indicated that each sentence can also constitute a highly independent block with its own independence. Thus, if both randomized texts at the sentence scale and texts with arbitrary punctuation all exhibit fractals, then it is difficult to consider fractals as an intrinsic feature of human writing, and studying literary style through fractals becomes almost meaningless.

## 4. Conclusion

An examination of the history of punctuation reveals that, although it greatly facilitates comprehension, it is not an intrinsic property of natural language. Modern Chinese punctuation, for example, was standardized in the early 20th century under Western influence, and a standardized system was largely absent from classical Chinese texts. Globally, punctuation emerged primarily as a reading aid—an auxiliary device to help convey meaning—rather than as a fundamental element in the generative process of language. Its current ubiquity reflects convention and standardization in written communication, not a universal necessity for linguistic expression.

The length of a sentence is a creative choice entirely at the writer's discretion. Unlike vocabulary selection, which is constrained by semantics, grammar, and context, sentence length offers considerable freedom for variation. It is a well-established fact that when the sentence-length sequence of a text is randomly shuffled by a computer, its long-range correlations are lost.[8] Building on this, the author of this study conducted a writing exercise in which a text was composed strictly following a sentence-length sequence obtained from such a computer-shuffled text. Initially, adhering to the randomized sentence-length sequence felt as rigid as composing traditional Chinese poetry under strict metrical rules; however, the process became progressively easier. While this exercise lacks the scientific rigor of a formal experiment, it anecdotally suggests that a text with a shuffled, uncorrelated sentence-length sequence can still yield naturally meaningful and coherent writing.

A key finding in our results is that while the experimental text (especially *Cao Ji*) exhibited a level of sentence independence nearly identical to that of the computer-shuffled text, its sentence-length sequence still retained long-range correlations. This observation prompts us to be cautious about treating long-range correlations as an inherent or indispensable feature of natural human language. Exaggerating their importance without sufficient empirical support could lead to conflating statistical regularities with the essential properties of linguistics, thereby creating a 'fractal illusion' in the interpretation of human writing behavior.

**Author contributions**

Ying Zeng and Junying Cui contributed equally to this work as co–first authors. Lejun Li also made important contributions.

**Acknowledgments**

We acknowledge that the data presented in this paper represent only a portion of the overall experiments. We extend our sincere gratitude to Maciej Kurzynski (Lingnan University, Hong Kong, China) for his exceptionally insightful feedback before the finalization of this study. We are also grateful to Guifu Yang (Tianjin, Jianhua Hospital, China) and Zhuoran Ma (University of Lyon III, France) for providing relevant materials. In addition, we thank the developers of random.org for supplying random text data. Finally, this study benefited from valuable discussions with numerous scholars in fractal research and digital humanities, whom we cannot list individually.

**Ethics Statement**

This study was approved by the School of Culture and Communication at Shandong University and was deemed exempt from full ethics review, as it involved only anonymous, non-sensitive data generated during regular classroom activities.

**Data Availability Statement**

The experimental data for this study are available from the corresponding author upon reasonable request.

**Conflict of Interest**

The authors declare that they have no conflict of interest.

**Note**

[1] For ease of comparison with the names of the experimental texts, we refer to the original Chinese title by its Pinyin transcription, *Fan Hua*.

[2] For ease of comparison with the original text, this translation strictly follows the sentence breaks and rhythm of the Chinese original, including all commas, periods, word divisions, and repetitions, without making any adjustments to the word order.

[3] In the experiment, participants randomly struck keys. With Chinese input, no deliberate selection was required: random keystrokes automatically generated Chinese characters, occasionally accompanied by a few English letters.

[4] ① *Six Sisters*, by Yi Bei; ② *Fortress Besieged*, by Qian Zhongshu; ③ *Wives and Concubines*, by Su Tong.

[5] The comparative tables are presented primarily for illustration, not for strict statistical comparison between literary and randomized corpora. Our central aim is to demonstrate that randomized texts, despite lacking semantic or narrative coherence, nevertheless exhibit long-range correlations in sentence length.

[6] The English translation of *Cao Ji* exhibits lower but still significant fractal characteristics, serving as an additional language dimension to validate the universal presence of fractal properties in random texts, rather than comparing the strength of fractals between Chinese (character-based) and English (word-based) analyses.

[7] In comparison with the works of shuffle literature, the similarity of *Cao Ji* lies exclusively in its nonlinear reading structure, rather than in any underlying literary purpose or aesthetic design.

[8] The experimental texts were also tested after computer-based shuffling, and the results likewise indicated that long-range correlations in sentence length had disappeared. Further details are provided in Appendix B.

**Appendix A.** The values of the original and shuffled sequences for two works of shuffle literature

Comparison of the two orders of *The Unfortunates*

| Text | $H_1$ | $H_2$ | $\beta_1$ | $\beta_2$ | $\Delta\alpha_1$ | $\Delta\alpha_2$ |
|---|---|---|---|---|---|---|
| The original order | 0.71 | 0.59 | 0.44 | 0.15 | 0.30 | 0.61 |
| The random order | 0.65 | 0.60 | 0.29 | 0.14 | 0.33 | 0.62 |

Comparison of the two orders of Chinese translation of *Composition No. 1*

| Text | $H_1$ | $H_2$ | $\beta_1$ | $\beta_2$ | $\Delta\alpha_1$ | $\Delta\alpha_2$ |
|---|---|---|---|---|---|---|
| The original order | 0.56 | 0.60 | 0.16 | 0.26 | 0.28 | 0.05 |
| The random order | 0.64 | 0.57 | 0.23 | 0.23 | 0.30 | 0.04 |

**Appendix B.** The Values of the Experimental Texts after Computer-based Shuffling

The Values of the Experimental Text *Hua* after Computer-based Shuffling

| Text | $H_1$ | $H_2$ | $\beta_1$ | $\beta_2$ | $\Delta\alpha_1$ | $\Delta\alpha_2$ |
|---|---|---|---|---|---|---|
| *Hua* | 0.49 | 0.49 | 0.02 | 0.02 | 0.37 | 0.13 |

The Values of the Experimental Text *Fan* after Computer-based Shuffling

| Text | $H_1$ | $H_2$ | $\beta_1$ | $\beta_2$ | $\Delta\alpha_1$ | $\Delta\alpha_2$ |
|---|---|---|---|---|---|---|
| *Fan* | 0.47 | 0.49 | 0.00 | 0.02 | 0.77 | 0.53 |

The Values of the Experimental Text *Jing* after Computer-based Shuffling

| Text | $H_1$ | $H_2$ | $\beta_1$ | $\beta_2$ | $\Delta\alpha_1$ | $\Delta\alpha_2$ |
|---|---|---|---|---|---|---|
| *Jing* | 0.53 | 0.51 | 0.00 | 0.02 | 1.04 | 0.93 |

The Values of the Experimental Text *Cao Ji* after Computer-based Shuffling

| Text | $H_1$ | $H_2$ | $\beta_1$ | $\beta_2$ | $\Delta\alpha_1$ | $\Delta\alpha_2$ |
|---|---|---|---|---|---|---|
| *Cao Ji* | 0.52 | 0.47 | 0.06 | 0.02 | 0.11 | 0.46 |

The Values of the Experimental Text English Version of *Cao Ji* after Computer-based Shuffling

| Text | $H_1$ | $H_2$ | $\beta_1$ | $\beta_2$ | $\Delta\alpha_1$ | $\Delta\alpha_2$ |
|---|---|---|---|---|---|---|
| English version of *Cao Ji* | 0.47 | 0.49 | 0.06 | 0.06 | 0.06 | 0.07 |

**Figures Legends**

**Figure 1.** Sentence length sequence comparison between *Hua* (black) and the original *Fan Hua* (red).

**Figure 2.** Sentence length sequence comparison between *Fan* (black) and the original *Fan Hua* (red).

**Figure 3.** Sentence length distribution in *Jing* with detected short and extremely long sentences.

**Figure 4.** Multiscale analysis of the sentence-length sequence of *Cao Ji*.

**Figure 5.** Sentence length distribution of *Cao Ji*, with detected short-sentence clusters and long-sentence spikes.